# Author's Post-Print (Final draft post-refereeing)

# An Image Classification Approach for Hole Detection in Wireless Sensor Networks

**Se-Hang Cheong · Kim-Hou Ng ·
Yain-Whar Si**

**Abstract** Hole detection is a crucial task for monitoring the status of wireless sensor networks (WSN) which often consist of low-capability sensors. Holes can form in WSNs due to the problems during placement of the sensors or power/hardware failure. In these situations, sensing or transmitting data could be effected and can interrupt the normal operation of the WSNs. It may also decrease the lifetime of the network and sensing coverage of the sensors. The problem of hole detection is especially challenging in WSNs since the exact location of the sensors are often unknown. In this paper, we propose a novel hole detection approach called FD-CNN which is based on Force-directed (FD) Algorithm and Convolutional Neural Network (CNN). In contrast to existing approaches, FD-CNN is a centralized approach and is able to detect holes from WSNs without relying on the information related to the location of the sensors. The proposed approach also alleviates the problem of high computational complexity in distributed approaches. The proposed approach accepts the network topology of a WSN as an input and generates the identity of the nodes surrounding each detected hole in the network as the final output. In the proposed approach, a FD algorithm is used to generate the layout of the wireless sensor networks followed by the identification of the holes in the layouts using a trained CNN model. In order to prepare labeled datasets for training the CNN model, an unsupervised pre-processing method is also proposed in this paper. After the holes are detected by the CNN model, two algorithms are proposed to identify of the regions of the holes and corresponding nodes surrounding the regions. Extensive experiments are conducted to evaluate the proposed approach based on different datasets. Experimental results show that FD-CNN can achieve 80% sensitivity and 93% specificity in less than 2 minutes.

Se-Hang Cheong
E-mail: dit.dhc@lostcity-studio.com

Kim-Hou Ng
E-mail: db62536@um.edu.mo

Yain-Whar Si
Faculty of Science and Technology, University of Macau, Macau, China
E-mail: fstasp@umac.mo

# 1 Introduction

Wireless sensor networks (WSN) have become one of the emerging technologies. They are applied to a wide range of application domains including environment and agricultural monitoring [6], object tracking and surveillance [32], traffic control [26], and disaster recovery [4] etc. A WSN often comprises of a number of tiny sensors (devices). The number of such sensors can vary from a few dozens to several thousands in which each sensor is capable of communicating with each other and they can perform limited data processing.

Designing efficient routing protocols and prolonging the network lifetime are crucial for maintaining WSNs. That is because sensors can communicate only with their neighbors. Sensors are usually installed with non-rechargeable batteries with limited capacity for sensing, data transfer, and data processing [1]. Formation of holes is considered to be one of the fundamental problems in WSNs. Holes may exist in a WSN caused by the errors in the sensor placement step or due to unavailable sensors caused by depleted or malfunction batteries [38]. When holes are formed, sensing or data transmission could be interrupted. The existence of holes in a WSN may reflect the network lifetime, sensing coverage, and energy consumption by the sensors [39].

Hole detection problem is one of the major interests in research community, especially for its wide range of applications in disaster recovery [34], monitoring scenarios [45], and underwater node coverage [27], etc. For example, the breakdown of a sensor area often indicates a critical event caused by an outbreak of a fire or destruction by an earthquake, etc. The problem of hole detection is especially challenging in WSNs which often contain low-capability sensors with unknown geographic location.

In recent years, a number of different approaches were proposed by researchers to solve the hole detection problem in WSNs. These approaches include topological algorithms [18] [7] [36], centralized approach [36], distributed approaches [19] [21], and transfer learning based approach [24]. However, these approaches are affected by issues such as high computation and communication cost, requirement on manual labelling of datasets, and inability to detect all different shapes of holes.

To this end, this paper proposes a novel approach called FD-CNN which is based on Force-directed (FD) Algorithm and Convolutional Neural Network (CNN) for detecting holes in location-free WSNs. The proposed approach takes advantage of both FD algorithms' capability in generating layouts from topology information and CNN's image recognition ability. By using state-of-the-art methods from information visualization and artificial intelligence domains, our approach can detect holes from WSNs based on a given network topology. To the best of authors' knowledge, the proposed approach is the first ever attempt to solve the hole detection problem in WSNs based on Force-directed Algorithms and Convolutional Neural Networks.

The main advantage of the proposed approach is that it does not rely on any location information (e.g. positions obtained from using Global Positioning System (GPS)) and sensors can be deployed either in a random or in a predefined manner.

In the proposed approach, a FD algorithm is used to generate potential layouts from the input network topology. Next, a trained CNN model is used to detect holes from these layouts. This process is iterated until the detection performance cannot be further improved. During the process of discovering a hole, our approach can also identify the identity (ID) of the sensors along the hole's boundary. Such detection allows the sensors located on the boundary of a hole to be cached locally in that region, whereby providing a conduit to improve the network routing and packet forwarding schemes.

Although CNNs are commonly used in object detection and face recognition tasks, they are rarely used for hole detection from graph layouts. In order to prepare labeled datasets for training CNN models, an unsupervised pre-processing method is also proposed in this paper. Extensive experiments are conducted to evaluate the proposed approach based on different datasets. Experimental results show that FD-CNN can achieve 80% sensitivity and 93% specificity in less than 2 minutes in sparse and uniform WSNs. The main contributions of this paper can be summarized as follows:

- A novel approach called FD-CNN which is based on FD Algorithm and CNN for hole detection in location-free WSNs is proposed in this paper.
- FD-CNN is able to detect holes from WSNs without relying on the information related to the location of the sensors.
- FD-CNN also alleviates the problem of high computational complexity which is often associated with distributed approaches.
- In order to prepare labeled datasets for training the CNN model, an unsupervised pre-processing method is also proposed in this paper.
- FD-CNN is evaluated by using 3 different FD algorithms when they are paired with a CNN for hole detection. Experimental results show that FD-CNN can achieve 80% sensitivity and 93% specificity in less than 2 minutes.

This article is organized as follows. Related work is briefly reviewed in Section 2. The proposed approach (FD-CNN) is discussed in Section 3. Experimental results of the proposed approach are described in Section 4. The conclusions and future works of the article are given in Section 5.

## 2 Related Work

In a WSN, sensing coverage reflects the quality of monitoring area by a sensor [35, 42]. Sensing coverage and conserving energy consumption are essential for improving the performance on data transfer and prolonging the network lifetime. Minimizing the energy consumption could prolong the lifetime of the sensors in maintaining the coverage in the WSNs. A hole can be considered as a region in a WSN where nodes are unavailable. Holes can cause interruption to the operation of a WSN since certain nodes cannot participate in normal communication activities [12]. Sensors are often deployed in a predefined manner called determined network coverage. However, in random network coverage, there is no prior information available about the location and topological structure of sensors [11].

Data in WSNs normally traverse multiples hop to reach the destination due to the short-range communication nature of the sensors. In recent studies, several schemes were proposed for the routing in WSNs with holes such as perimeter routing [44] and geographic routing [30]. In the perimeter routing, data packets are forwarded along the boundary of a hole. However, routing paths are enlarged and traffic is concentrated along sensors holes' boundary. The idea of geographic routing is to create a region covering the hole in which all the packets stay away from these regions. However, packets in geographic routing may be stopped at the hole boundaries because packets are not allowed to be forwarded along sensors holes' boundary [2]. In addition, a source node needs to know the location of the destination node in geographic routing, either by acquiring it from the Global Positioning System (GPS) or similar location services.

Four types of holes are reported in recent studies. They are coverage hole, routing hole, sink/wormhole, and jamming hole [29]. If there are regions in a WSN which are not covered by any sensor, then coverage holes may have occurred. If a routing hole exists in the deployed WSN, then sensors may not be able to communicate with each other. In the event of radio frequency is furnished with jammers, then jamming holes will occur. Sensors cannot send/receive data in sink/wormholes due to malefic nodes blocking the transmission during denial of service attacks.

Several works have addressed the issue of coverage hole detection in WSNs. Fang et al. [16] proposed an approach for coverage holes detection based on the location of the sensors. The authors also proposed an algorithm which sweeps along hole boundaries in order to discover nodes along the boundary of holes by using the geographical location of sensors.

An approach for coverage hole detection and sensor deployment based on a FD algorithm is proposed in [40]. In this approach, sensors located at an appropriate distance from the hole are first detected. Then the location of the sensor is determined by the FD algorithm.

Topological hole detection algorithms are based on the network topology of the WSNs in which only local connectivity information is available [18]. Bi et al. [7] proposed a converge hole detection algorithm which is based on an assumption that number of neighbors of a sensor locating on the boundary of a hole is less than others. In their approach, a sensor located on the boundary of a hole is classified by its degree and the average degree of its 2-hop neighbors. Each node determines whether it is on the boundary of a hole by comparing the difference of the average degree with its 1-hop and 2-hop neighbors. Specifically, the proposed method [7] counts the number of neighbors at each sensor and classify them when the number of neighbors is less than a threshold. However, the proposed algorithm involves a huge communication overhead [5]. Besides, their approach is also inefficient for detecting holes from a large WSN and not all nodes on the boundary of hole can be identified correctly [40].

In [36], Ramazani et al. proposed a centralized coverage hole detection algorithm to construct a coverage planar graph of a WSN in which location of the nodes are estimated using the received signal strength. A plane simplicial complex is built in the proposed algorithm to identify the sensors from holes' boundaries of the coverage planar graph. The proposed FD-CNN approach is purely based on the information of the network topologies which is available when the sensor network is deployed at the very beginning and no communication is needed during the runtime to retrieve the topology information. Yan et al. also proposed an approach for hole detection based on simplicial complex reduction algorithm [43] in which redundant sensors are deployed to achieve $k$-coverage of WSNs, where $k$ is the number of sensors that can cover a monitoring region. However, the approach proposed by [43] has high computational complexity because it needs to construct a complex structure of the network for hole detection. In the proposed FD-CNN approach, a FD algorithm is used to produce the layout of WSNs without rely on anchor and sink sensors.

In recent years, increasing number of studies adopted CNNs for WSNs related applications. CNN is a class of deep neural network from Artificial Intelligence area. CNNs can detect and recognize general patterns found in the input images [33]. Recent studies also reported that CNNs can be useful for object detection task such as localization of the objects from remote sensing images. In [25], the authors adopt CNNs to detect and locate the objects (e.g., aircrafts, oil tanks, vehicles, etc.) from aerial images. Tong et al. [41] also proposed a CNN based approach to improve the accuracy of event classification in homogeneous sensor networks. In [3], Ahn et al. proposed a post-processing approach in which a CNN was executed at the back-end server and a WSN was deployed for bird nest monitoring. The objective of the proposed approach by Ahn et al. is to reduce the size of images to be transferred while monitoring the WSN. In their work, a CNN is used to maintain the classification accuracy of the birds from highly compressed sensor images. In [19], Hajjej et al. proposed a distributed reinforcement learning approach for WSNs where the nodes can recover from coverage constraints through local acquaintances, such as adjusting the sensing range and locations of sensors based on the detected holes. Khalifa et al. [21] also proposed a distributed hole detection and restoration of nodes which are previously positioned in the sensor network. Frequent communication and transmission will be needed for detecting the holes in distributed approaches. The proposed FD-CNN approach does not involve any communication and data exchange when hole detection is performed in wireless sensor networks since only the network topology is used for calculation. Besides, Lai et al. [24] proposed a transfer learning approach for coverage hole detection. The main focus of the approach in [24] is to detect triangular coverage holes. However, the proposed approach by Lai et al. is a supervised approach and it requires labelling of the dataset manually. The proposed FD-CNN approach is an unsupervised approach which does not require manual labelling of training datasets. Our approach detects non-triangular shaped coverage holes.

In our work, we focus on the detection of converge hole in location-free WSNs. The proposed approach assumes random network coverage in which sensors are randomly deployed in the target area. Besides, our approach is also applicable for determined network coverage and can be used to detect arbitrary shape of holes. In addition, the proposed FD-CNN is an unsupervised approach which does not require manual labelling of training datasets. Besides, FD-CNN is purely based on the information of the network topologies and it does not rely on local or global positioning systems. In the proposed approach, a FD algorithm is first used to produce the layout of WSNs. Next, a CNN is used to detect the holes from the generated layout. To better differentiate the proposed approach and the existing methods reviewed in this paper, a brief comparison of the features is given in Table 1.

Table 1: Comparison of the proposed approach and the existing methods reviewed in this paper.

| Methods | Brief Summary |
|---|---|
| Bi et al. [7] | – It is assumed that number of neighbors of a sensor locating on the boundary of a hole is less than the number of neighbors of a normal sensor node.<br>– High communication overhead may be involved for synchronizing the neighborhood information. |
| Ramazani et al. [36] | – A centralized coverage hole detection algorithm is used to construct the coverage planar graph of WSNs.<br>– Location of the nodes are estimated using received signal strength. |
| Yan et al. [43] | – Based on simplicial complex reduction algorithm.<br>– A complex structure of simplicial complex is constructed for hole detection. |
| Hajjej et al. [19] | – A distributed approach based on reinforcement learning.<br>– Communication overhead exists in sensors deployed in the WSN. |
| Khalifa et al. [21] | – A distributed self-healing algorithm for hole detection.<br>– It utilizes the information of nodes previously positioned in the sensor network. |
| Lai et al. [24] | – Based on force-directed algorithms and transfer learning.<br>– A supervised approach and detect triangular coverage holes.<br>– Require manual labelling of the dataset. |
| Our Approach | – Based on force-directed algorithms and CNN.<br>– Information on location of the sensors is not needed for computation and only requires the network topology of the WSN as an input.<br>– Unsupervised approach and can detect non-triangular coverage holes. |

## 3 Hole Detection in WSN with FD and CNN

The overview of using a FD Algorithm and a CNN for detecting coverage holes in location-free WSNs is described in this section. An example of holes in a WSN is depicted in Figure 1. In this example, the network contains 14 nodes and 21 edges. The region of the holes are colored in purple. The nodes located on a hole

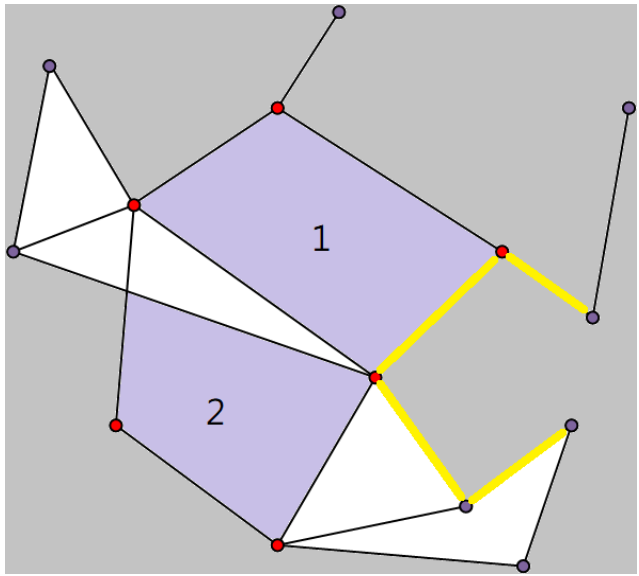

Fig. 1: An example of holes in a WSN.

are shown in red color. In this example, there are two holes which are labeled as 1 and 2.

In our approach, we target the inner holes in a WSN for detection. As illustrated in yellow lines in Figure 1, outer (outward) holes are not considered in this work because they are actually the boundary of WSNs. Before the overall design of the proposed FD-CNN approach is explained in detail, the characteristics of the sensors, connections, and the type of holes considered in our framework can be summarized as follows:

– The connection range of sensors in the WSNs considered in our approach is homogeneous.
– A hole must be a closed area and surrounded by nodes (sensors).
– A hole is an irregular shape composed of nodes and edges.
– A hole must be surrounded by at least 4 nodes. holes formed by 3 nodes are not considered since nodes which forms a triangular shape are able to communicate with each other due to their close proximity (connected each other in a 1-hop distance within a homogeneous connection range). Therefore, white regions in Figure 1 are not considered as holes in our work.

3.1 Overview

A high-level overview of the FD-CNN approach is illustrated in Figure 2. The input to the FD-CNN is a network topology of a WSN and the output is the identity (ID) of the nodes surrounding each detected hole in the network. The pseudo code for the proposed FD-CNN approach is given in Algorithm 1. First, a FD algorithm is used to generate the layout (image) of the WSN. Next, a CNN model is used to detect converge holes from the layout. Once holes are detected by the CNN model, two algorithms are used to perform post recognition tasks which are described in Section 3.3.5. These tasks include identification of the regions of the holes from the layouts and recognition of nodes located along the boundary of the regions.

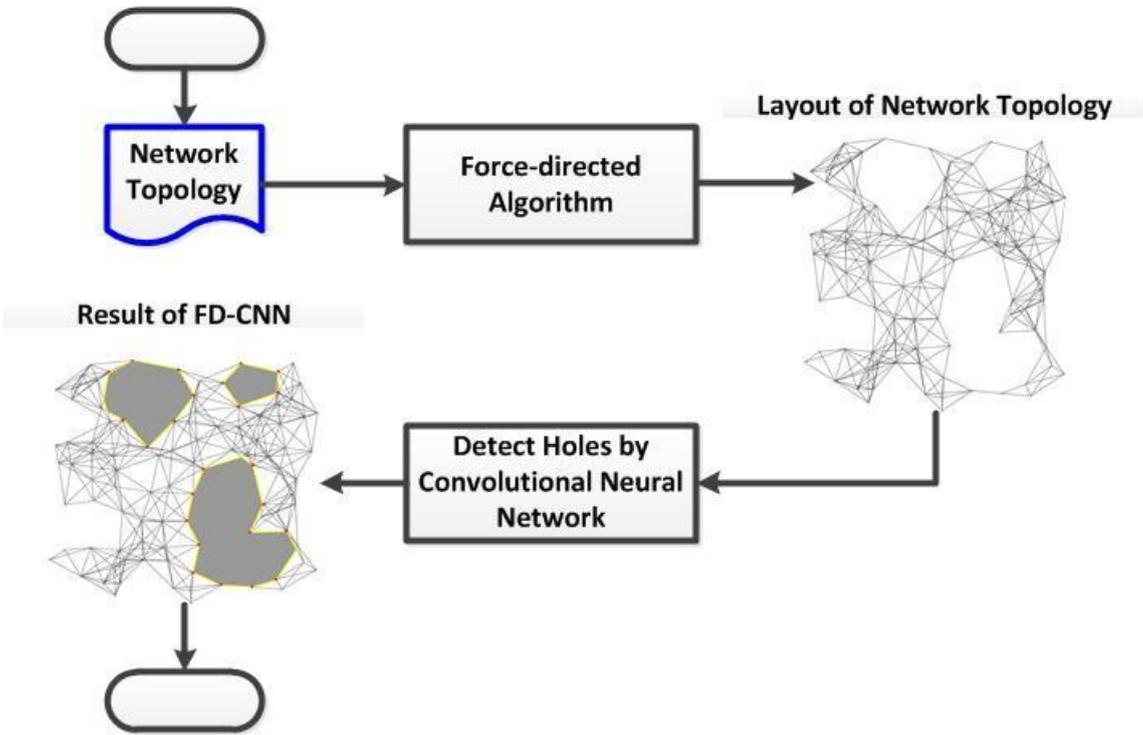

Fig. 2: Overview of FD-CNN approach.

---

**ALGORITHM 1: (FD-CNN)** Pseudo code of FD-CNN.

**Input:** A text file *dataset* containing the network topology of WSN.
**Output:** List of nodes along the holes' boundaries

// Initialize variables.
initialize a layout container *layout* storing the layout information of WSN;

// 1) Generate a new layout of WSN by the FD algorithm.
*layout* = **fd**(*layout*);

// 2) Detect holes with CNN.
*boxes* = **YOLOv3-Darknet**(*layout*);

// 3) Identify holes' boundaries given in Algorithm 2.
*holes* = **HoleIdentify**(*boxes*);

// 4) Identify nodes along the holes' boundaries given in Algorithm 3.
*bnodes* = **SensorIdentify**(*holes*);
**return** bnodes;

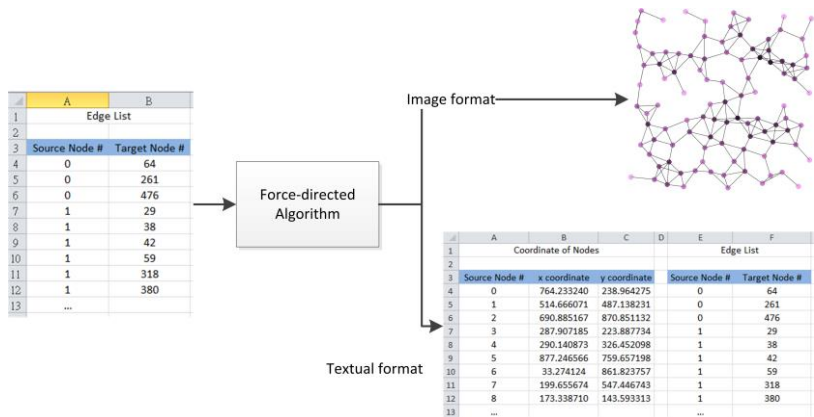

Fig. 3: The input and output of FD algorithms.

3.2 Force-Directed Algorithm

A Force-directed (FD) Algorithm is a kind of graph drawing algorithm which can be used to generate the visualization of the graphs by using the information contained in the network topology. There are several models of FD algorithms, such as accumulated force model, energy function minimization model, and combinatorial optimization model [10]. A simplified illustration of a FD algorithm is depicted in Figure 3.

The input to a FD algorithm is a set of nodes and the edges connecting these nodes. The set of nodes and edges represent the network topology of a WSN. The output of a FD algorithm is the layout of the network topology. A layout can be considered as a possible snapshot of the WSN. Due to the variation in force calculation, for a given input (topology), different layouts will be generated by each FD algorithm at a specific point in time. Moreover, in majority of the FD algorithms, the layouts generated from the input topology can be exported into an image or a text format file for post-processing. This option is illustrated in Figure 3.

In the textual output, the projected *x* and *y* coordinates of each node is recorded in the file. In our work, three FD algorithms were adopted for evaluation. They are Davidson Harel (DH) algorithm [13], ForceAtlas2 (FA2) algorithm [20], and Kamada Kawai with Multiple Node Selection and Decaying Stiffness (KK-MS-DS) algorithm [8]. Moreover, the force models, characteristics, and time complexity of these FD algorithms are summarized in Table 2.

*Davidson-Harel algorithm:* Davidson Harel (DH) algorithm [13] uses the simulated annealing to minimize edge crossings. The DH algorithm also prevents nodes from moving too close to non-adjacent edges. Simulated annealing is inspired by the physical cooling process of the molten material. If the molten steel cools too quickly, it will burst and form bubbles, making it brittle. Therefore, in order to obtain better results, the steel must be cooled uniformly in a process called an-nealing in metallurgy [22]. For the detailed formulation of DH algorithm, pleaser refer to Appendix A and Appendix B.

Table 2: Comparison of adopted FD algorithms for evaluation.

| FD algorithm | Force Model | Time Complexity | Characteristics |
|---|---|---|---|
| FA2 | Accumulated Force Model | $O(V^2 + E)$ | Repulsive and attractive forces are used in the accumulated force models. |
| KK-MS-DS | Energy Function Minimization Model | $O(n \times (V + v^2))$ | Attractive and repulsive forces are not considered separately, but rather used in conjunction to minimize the energy function. |
| DH | Combinatorial Optimization Model | $O(V^2 \times E)$ | Simulated annealing technique is used to find local minima of the energy function. |

*ForceAtlas2 algorithm:* The objective of ForceAtlas2 (FA2) algorithm [20] is to meet the speed and accuracy requirements of network visualization. The FA2 algorithm extends LinLog [31] and FR algorithm [17]. Jacomy et al. [20] also proposed a revised attractiveness based on the LinLog model. For the detailed formulation of FA2 algorithm, please refer to Appendix A and Appendix B.

*KK-MS-DS algorithm:* KK-MS-DS algorithm [8] is designed to push the nodes which are located along the outer boundary away from the inner nodes. In KK-MS-DS algorithm, nodes are tagged with a decaying stiffness. The higher the decaying stiffness value of the node, the farther the distance of the node can be moved. For the detailed formulation of KK-MS-DS algorithm, please refer to Appendix A and Appendix B.

3.3 Convolutional Neural Network (CNN)

CNNs are widely used for analyzing images. A typical CNN composes of convolutional and dense layers [28]. The dense layer learns global patterns from input images, while the convolutional layer extracts the general pattern found in small windows of the input image. Object detection aims at simulating the human visual system for detecting pixels or regions of holes.

3.3.1 CNN Models

Since CNN models have never been used for detecting holes from the images of layouts representing WSNs, there are currently no datasets available in public domains. In order to prepare the labeled datasets for training CNN models, a unsupervised pre-processing method is proposed in this paper. The general overview of the proposed pre-processing method is illustrated in Figure 4. In this method, the layout of WSN which is generated by a FD algorithm from the previous step is first converted into a bitmap image. Next, the region (rectangle area) of holes are labeled automatically in the bitmap image. Next, the centroid of labeled holes is calculated. Finally, the region (rectangle area) and centroid of labeled holes are stored as a Hole Detection Training Object which will be used for the training of the CNN model. Specifically, the objective of the pre-processing method is to create images that include holes for recognitions. These labeled holes conform to the relevant definition of a hole stated at the beginning of section 3. The proposed pre-processing method includes three steps: (1) Graph Segmentation; (2) Redundant Edge Reduction, and (3) *K*-node Hole.

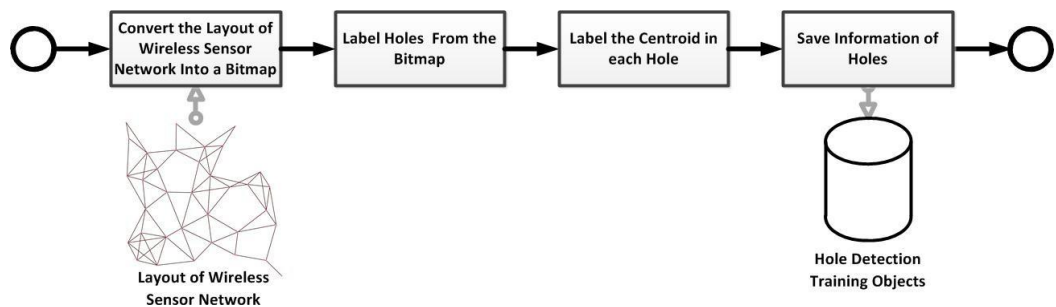

Fig. 4: The overview of the proposed pre-processing method for preparing the training datasets for CNN.

3.3.2 Graph Segmentation

In this step, holes are cut out in pieces from the layout (bitmap image) of a WSN generated by the FD algorithm. Holes are then segmented from the cutouts as illustrated in Figure 5(a). Next, each extracted image is saved into a bitmap image in the training dataset.

3.3.3 Redundant Edge Reduction

Existence of redundant edges is a major weakness for Graph Segmentation step because holes are cut out from an image directly. Too much irrelevant edges in the cutouts can increase noise in extracted features. In this paper, an approach called *Redundant Edge Reduction* is proposed to remove redundant edges outside the region of the holes. It can be achieved by parsing the formatted text output of the layout of WSNs instead of using the bitmap outputs. Specifically, the position of the nodes which are located on the boundary of the holes can be parsed using the textual outputs of the layouts generated by FD the algorithms[1]. Bitmap images of holes are then recreated from scratch based on the positions (coordinates) information without redundant edges. The resolution and color of the nodes and edges are also modified in the process of Redundant Edge Reduction.

---

[1] FD algorithms can generate both textual and graphical outputs from a given input topology. Textual output contains the coordinates of the nodes in the projected layouts.

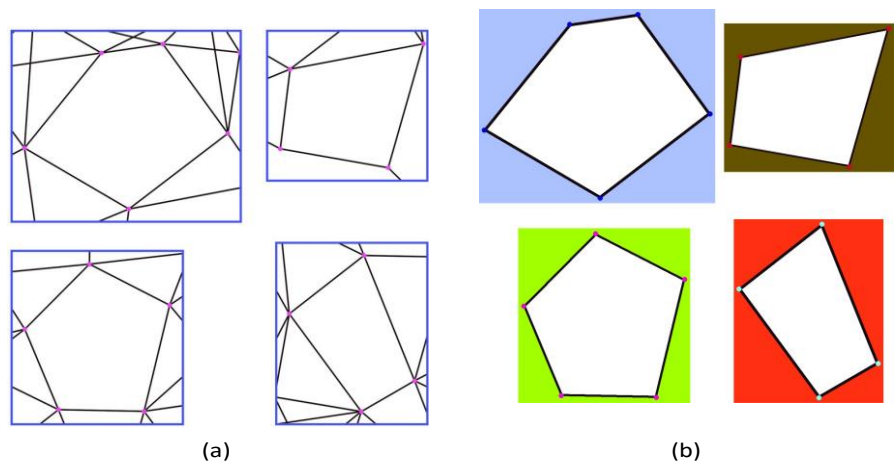

Fig. 5: (a) Extracted images by using Graph Segmentation. (b) Result of Redundant Edge Reduction step.

*3.3.4 K-node Hole*

Features extracted from Graph Segmentation and Redundant Edge Reduction steps do not contain the information of nodes located on the boundary of holes, which can be a valuable source of information for classification. To improve this situation, a step called *K-node Hole* was introduced in this paper. The aim of the proposed step is to categorize holes from labeled images that are encircled with the number of boundary nodes. For example, holes with 4 boundary nodes are defined as 4-node Hole, holes with 5 boundary nodes are defined as 5-node Hole, and so on. In addition, holes with 7 or more boundary nodes can enclose different shapes forming polygons. Therefore, in this paper, *k*-node Hole for holes with 7 or more boundary nodes is defined as illustrated in right-bottom corner of Figure 6.

Once the training datasets are prepared by the proposed pre-processing method, these data sets are used for the training of the CNN models. The training process of the CNN model is executed iteratively and the accuracy and average loss are calculated at each iteration. The training terminates when the average loss is less than a predefined threshold. Finally, the trained model of CNN is stored for hole detection.

*3.3.5 Hole Detection with CNN*

After a CNN model is trained with the labeled data sets, it can be applied for detecting holes from the layouts generated by the FD algorithm. The overview of hole detection by the trained CNN model is illustrated in Figure 7. In the hole detection process, the layout of the WSN is first generated by using a FD algorithm. Next, the trained CNN model is used to detect the holes from the layout of WSN. In our work, YOLOv3 [37] was adopted for constructing the CNN models. YOLOv3 proposed by Joseph Redmon [37] is an object detection algorithm based on CNN. YOLOv3 is able to achieve high accuracy without the need of high performance computing hardware. The architecture of YOLOv3 is depicted in Figure 8. YOLOv3 uses a Darknet variant called Darknet-53 [37] which has 53 convolution layers trained on Imagenet [15, 23]. In YOLOv3, 53 layers were stacked on top of Darknet-53, thus providing 106 layers of fully convolutional underlying architecture. This is the reason why YOLOv3 greatly improves the detection accuracy compared to YOLOv2.

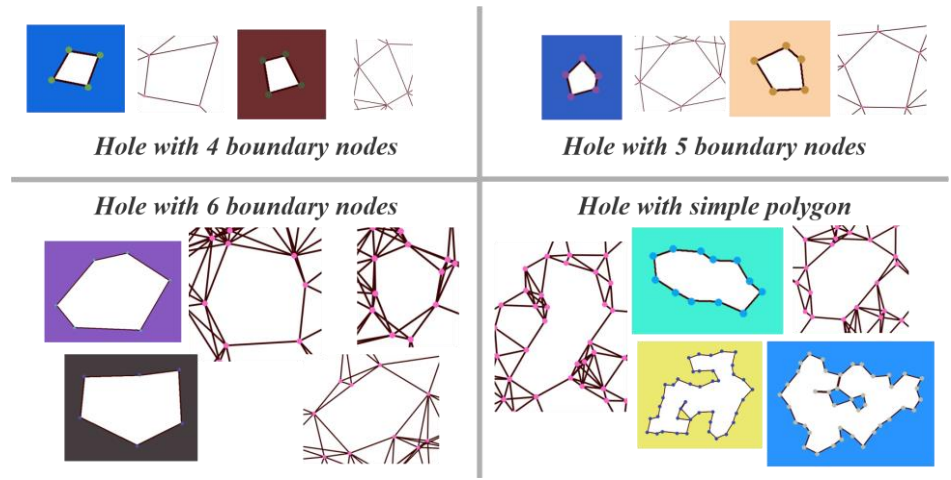

Fig. 6: Classification of holes by *K*-node Hole.

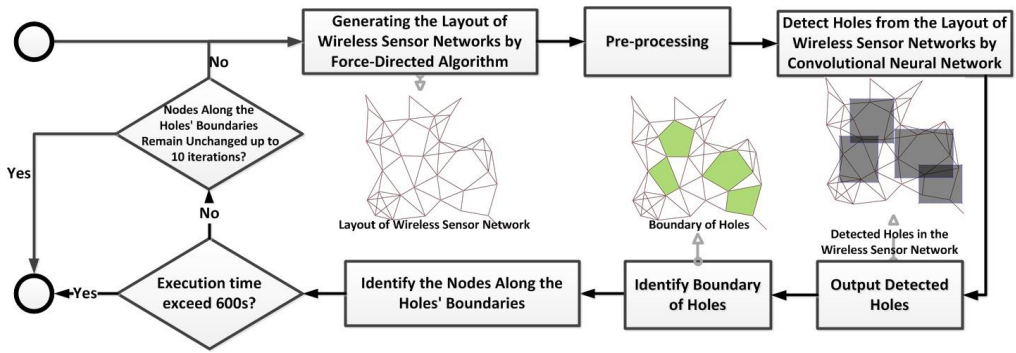

Fig. 7: The overview of hole detection in Wireless Sensor Networks based on Force-directed (FD) Algorithm and Convolutional Neural Network (CNN)

After the holes are detected by the CNN model, two algorithms are used to perform post recognition tasks. Specifically, these tasks include identification of the region of the holes from the layouts and recognition of the node IDs located on the boundary. The psuedocode for identifying the nodes along the holes' boundaries is given in Algorithm 2. The algorithm fills the regions of detected holes with green color. In this algorithm, a flood fill method is applied at the centroid of holes to identify the regions. In this example, the regions of predicted holes are filled with green color. The input of proposed algorithm contains the network topology of

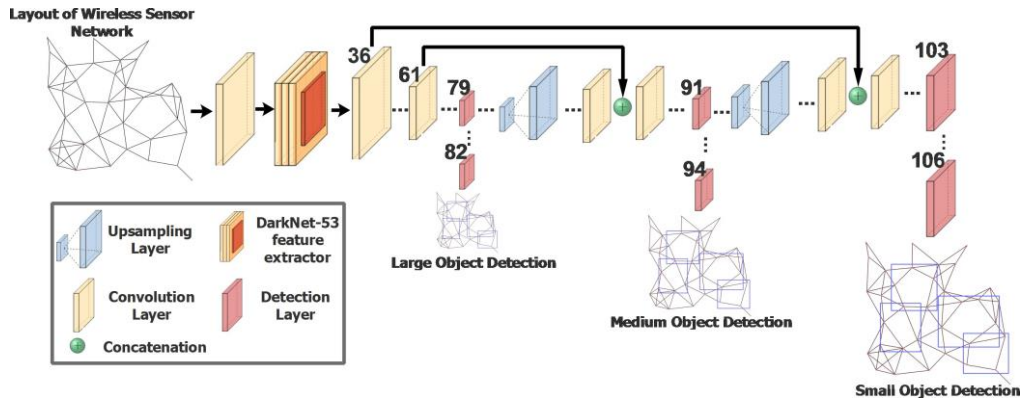

Fig. 8: The architecture of YOLOv3.

WSN (*G*) and the centroid of the detected holes (*c*). The algorithm has five major steps:

1. A bitmap image of filled holes is initialized.
2. Draw nodes on the bitmap image.
3. Draw edges on the bitmap image.
4. Fill the region of holes starting from the centroid of the detected hole.
5. Return the bitmap image of filled holes.

---

**ALGORITHM 2: (HoleIdentify)** Algorithm for identification of holes' boundaries.

**Input:** Network topology of WSN (*G*), and centroid of the detected hole (*c*).
**Output:** a bitmap image of filled holes (*b*).

let *H* and *W* be the height and width of the bitmap image;
let *V* be the set of nodes in *G*;
let *E* be the set of edges in *G*;
initialize a bitmap *b* storing the layout of filled holes.

let $color_t$ be the color to be replaced;
let $color_r$ be the color of the filled region of holes;
let *Q* be the queue containing position to be checked in bitmap image *b*;
$color_t$ = *white*;
$color_r$ = *green*;
$Q \leftarrow \{c\}$;
**repeat**
  *n* = *dequeue*(*Q*);
  **if** *the color of the position m to the north, south, east or west of n is $color_t$* **then**
    Fill the color of bitmap image *b* at position *m* to $color_r$;
    $Q \leftarrow \{m\}$;
  **end**
**until** *Q is empty*;
**return** *b*.

---

Once the region of holes are determined, a contour tracing algorithm is used to identify the nodes located on the boundary of holes. The psuedo code for iden-tifying the nodes along the holes' boundaries is given in Algorithm 3. The input of the algorithm contains the network topology of WSN (*G*) and the bitmap of filled image (*b*) which are produced from Algorithm 2. The Algorithm 3 has 4 steps:

1. Find contours from the bitmap of filled image.
2. Test if intersection exists in the node and contours.
3. Save nodes along the holes' boundaries.
4. Return the list of nodes from the holes' boundaries.

**ALGORITHM 3: (SensorIdentify)** Algorithm for identification of nodes along the holes' boundaries.

**Input:** Bitmap of filled image (*b*), and network topology of WSN (*G*).
**Output:** Nodes along the holes' boundaries (*bnodes*).

let *V* be the set of nodes in *G*;
let $color_T$ be the color of the filled region of holes;
$color_T$ = *green*;
*bnodes* = {};
let *contours* be the list of contours found from the bitmap of filled image.;

**for** *each contour c in contours* **do**
    **for** *each node v in V* **do**
        **if** *pointPolygonTest(v, contour) = TRUE* **then**
            *bnodes* ← {*v*};
        **end**
    **end**
**end**
**return** *bnodes*.

The process of hole detection terminates once the execution time exceeds $T_s$ seconds or the node IDs located on the boundary of holes remain unchanged up to $T_i$ iterations. The detection output of the proposed CNN model contains the centroid and the area of predicted holes.

## 4 Experiments

### 4.1 Experiment settings

In this experiments, the total number of nodes *n* is set to 1,000, 2,000 and 3,000. The average degrees of nodes *d* is set to 6, 8 and 10 for generating topologies for the datasets. We evaluate the proposed approach on two types of datasets which includes sparse and uniform networks. In total, 18 WSNs were generated for the experiments based on all possible combinations of *n* and *d*. Moreover, the generation of datasets were based on [8] and they were generated by using CNCAH Network Generator [9].

In the proposed FD-CNN, FD algorithms are used to produce the layouts (images) of the WSNs from input topologies. For a given input topology, the quality of the layout generated by different FD algorithm could vary significantly. An example of such situation is illustrated in Figure 9(a) and Figure 9(b). In these figures, both layouts are generated by using two different FD algorithms with the same input topology.

Therefore, three FD algorithms (DH, FA2, and KK-MS-DS) were selected for evaluation in our experiments. To obtain a fair comparison in our experiments, each FD algorithm is implemented to stop the execution either when the maximum execution time exceeds $T_s$ = 600 seconds or the node IDs located on the boundary of holes remain unchanged up to $T_i$ = 10 iterations. For example, if the node IDs located on the boundary of holes from the FD algorithm remain unchanged since $100^{th}$ iteration, then the FD algorithm terminates at the $110^{th}$ iteration. Therefore, the layout of WSN generated at $100^{th}$ iteration will be considered as the final layout.

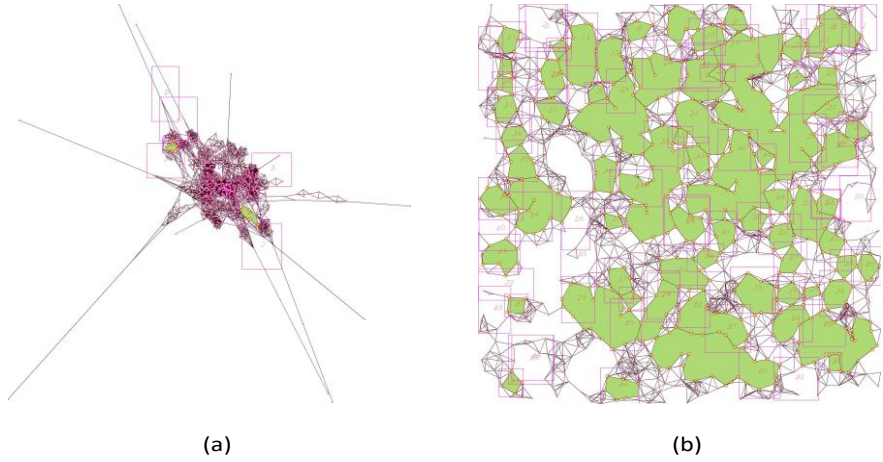

(a)  (b)

Fig. 9: Two example layouts generated by two different force-directed algorithms for the same topology. The detected holes are marked in pink color and the region of a hole is filled in green color. (a) The generated layout in which nodes are stacked on each other, (b) The generated layout in which nodes are placed uniformly.

Since the numbers of true positive $Hole_N$ and true negative $Hole_P$ in our datasets vary greatly, *accuracy* cannot reflect the real performance of algorithms when they are evaluated for hole detection. To alleviate this problem, *sensitivity* and *specificity* value were used as performance evaluation criteria.

4.2 Experiment results for sparse WSNs

The sensitivity and specificity of FD-CNN in sparse WSNs from the layout generated by KK-MS-DS, FA2 and DH algorithms were analyzed in this section.

*4.2.1 Dataset with 1,000 nodes*

Sensitivity of FD-CNN for different average degree is depicted in Figure 10. In this experiment, the total number of nodes was set to 1,000 and networks with average degree 6, 8 and 10 were tested. The experimental results reveal that the proposed algorithm achieves highest sensitivity when KK-MS-DS algorithm was used. Specifically, sensitivity is above 80% when the average degree $d$ = 8 and $d$ = 10 were used in the testing (see Figure 10(b) and Figure 10(c)). The highest sensitivity is around 50% when FA2 algorithm was used (see Figure 10(b)). The proposed algorithm achieves lowest sensitivity when DH algorithm was used that the sensitivity is lower than 10%.

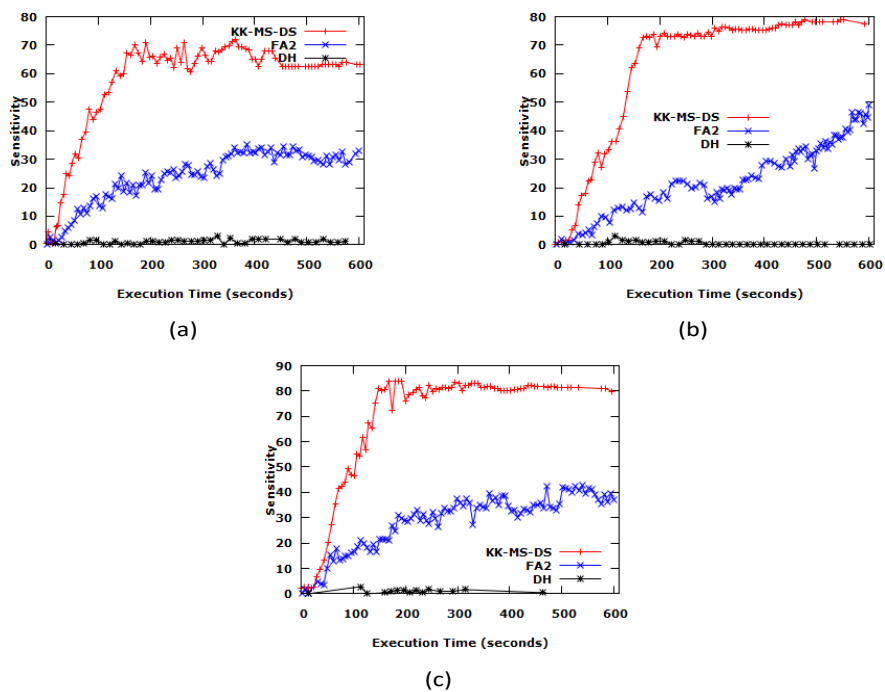

Fig. 10: Sensitivity when $n = 1,000$ and (a) $d = 6$, (b) $d = 8$, (c) $d = 10$ in WSNs.

Specificity of FD-CNN for different average degree is depicted in Figure 11. In this experiment, average degree 6, 8 and 10 are tested. The experimental results reveal that the proposed algorithm achieved the highest specificity (approx. 90%) when the average degree $d = 10$ is used in KK-MS-DS algorithm (see Figure 11(c)). However, the proposed algorithm produced relatively poor specificity when the average degree is low (i.e. when $d = 6$ and $d = 8$, see Figure 11(a) and Figure 11(b)).

The highest specificity of the proposed algorithm is approximately 80% when FA2 algorithm was used. The process of hole detection for DH algorithm was terminated at around $480^{th}$ second because node IDs located on the boundary of holes remain unchanged up to 10 detections. This situation can be observed from Figure 11(c).

*4.2.2 Dataset with 2,000 nodes*

Sensitivity of FD-CNN for different average degree is depicted in Figure 12. In this experiment, average degree 6, 8 and 10 are tested. For the layouts generated by the KK-MS-DS algorithm, the results showed that the proposed algorithm achieved the highest sensitivity despite the differences in the average degree. The sensitivity was higher than 50% when the execution time exceed 250 seconds (see Figure 12(b)). The proposed algorithm produced similar sensitivity for the layouts generated by the FA2 and DH algorithms.

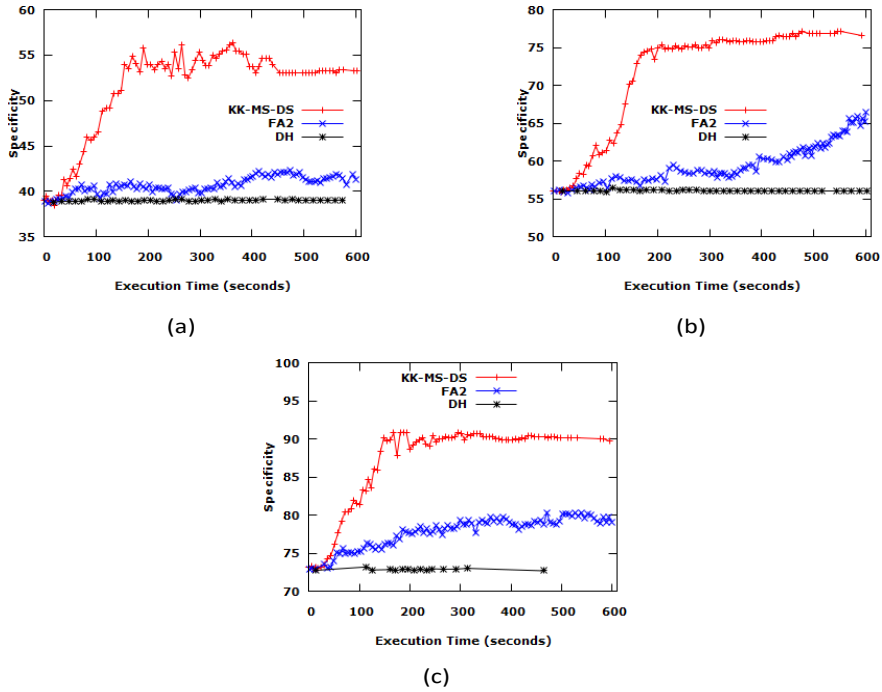

Fig. 11: Specificity when *n* = 1,000 and (a) *d* = 6, (b) *d* = 8, (c) *d* = 10 in WSNs.

Specificity of FD-CNN for different average degree is depicted in Figure 13. In this experiment, average degree 6, 8 and 10 are evaluated. The experimental results revealed that the proposed algorithm achieved the highest specificity with the average degree *d* = 8 and *d* = 10 as depicted in Figure 13(b) and Figure 13(c). Furthermore, the proposed algorithm produced similar specificity from the layout generated by the FA2 and DH algorithms regardless of the average degree. According to the experimental results depicted in Figure 13(a) and Figure 13(c), the specificity of proposed algorithm with a high average degree (*d* = 10) was better than a low average degree (*d* = 6).

### 4.2.3 Dataset with 3,000 nodes

Sensitivity of FD-CNN for different average degree is depicted in Figure 14. In this experiment, average degree 6, 8 and 10 are tested. Figure 14(b) showed that the proposed algorithm achieved the best sensitivity result for KK-MS-DS when the average degree *d* = 8 in used in WSNs. The proposed algorithm produced low

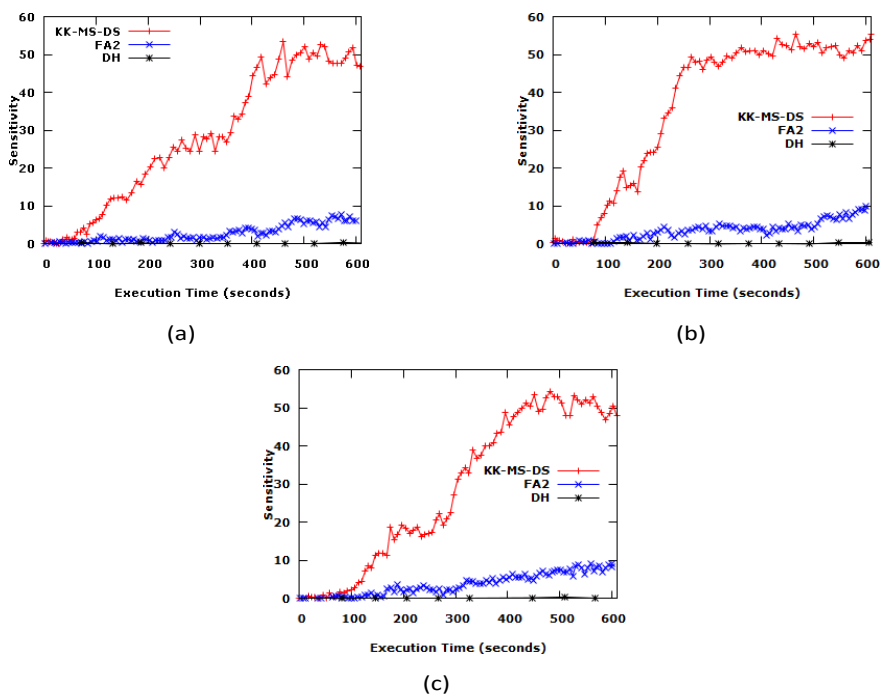

Fig. 12: Sensitivity when $n = 2,000$ and (a) $d = 6$, (b) $d = 8$, (c) $d = 10$ in WSNs.

sensitivity (i.e., lower than 10%) from the layout generated by the FA2 and DH algorithms.

Specificity of FD-CNN for different average degree is depicted in Figure 15. In this experiment, average degree 6, 8 and 10 are tested. From the results, we found that KK-MS-DS achieved the best specificity among three algorithms.

*4.2.4 Summary*

All in all, according to the experimental results, FD-CNN achieves the highest sensitivity and specificity from the layouts generated from the KK-MS-DS algorithm. When average degree is set to 6, sensitivity of FD-CNN is above 80%. In addition, the sensitivity and specificity of FD-CNN from the layout generated from FA2 algorithm is superior than DH algorithm especially when $n = 1000$. We also found that the performance of FD-CNN is unstable from the layouts generated by the DH algorithm. It achieves lowest sensitivity and specificity in the experiments.

4.3 Experiment results for uniform WSNs

The sensitivity and specificity of FD-CNN for uniform WSNs were evaluated in this section.

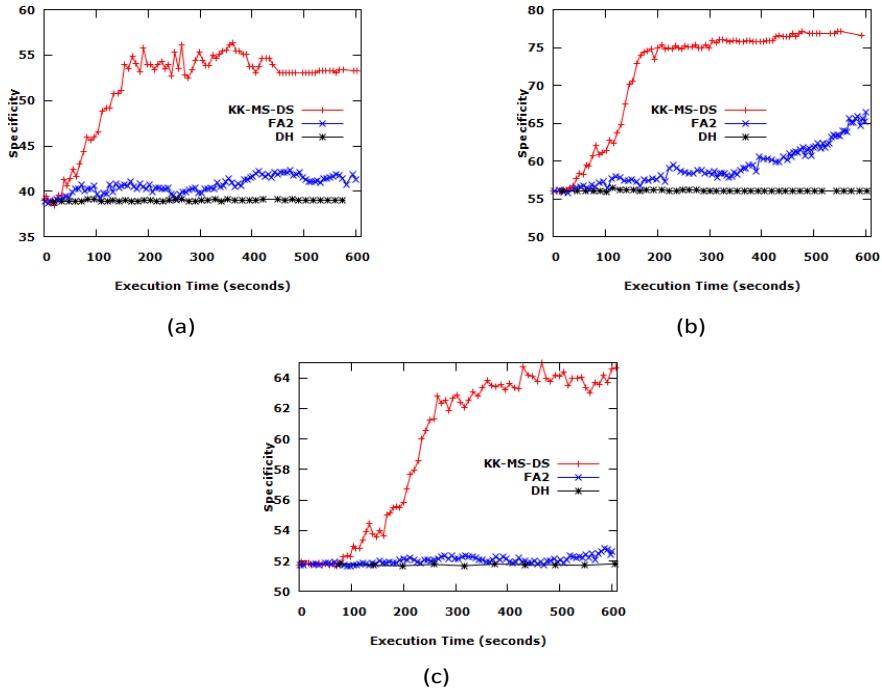

Fig. 13: Specificity when *n* = 2,000 and (a) *d* = 6, (b) *d* = 8, (c) *d* = 10 in WSNs.

*4.3.1 Dataset with 1,000 nodes*

Sensitivity of FD-CNN for different average degree is depicted in Figure 16. In this experiment, the total number of nodes is set to 1,000 and average degree 6, 8 and 10 are tested. The experimental results revealed that the proposed algorithm achieved the highest sensitivity with average degree *d* = 8 in uniform sensor networks for KK-MS-DS (see Figure 16(b)). The detection for FA2, KK-MS-DS and DH algorithms stops before time limit (i.e. 600 seconds) since the node IDs located on the boundary of holes was unchanged for more than 10 iterations as illustrated in Figure 16(c).

Specificity of FD-CNN for different average degree is depicted in Figure 17. In this experiment, the total number of nodes is 1,000 and average degree 6, 8 and 10 are tested. The highest and lowest specificity of proposed algorithm were obtained from the layouts generated from KK-MS-DS and DH algorithms regardless of the average degree. The highest specificity was around 94% when KK-MS-DS algorithm was used. The results also showed that the proposed algorithm achieved the highest specificity when the average degree was high (see Figure 17(b) and Figure 17(c)).

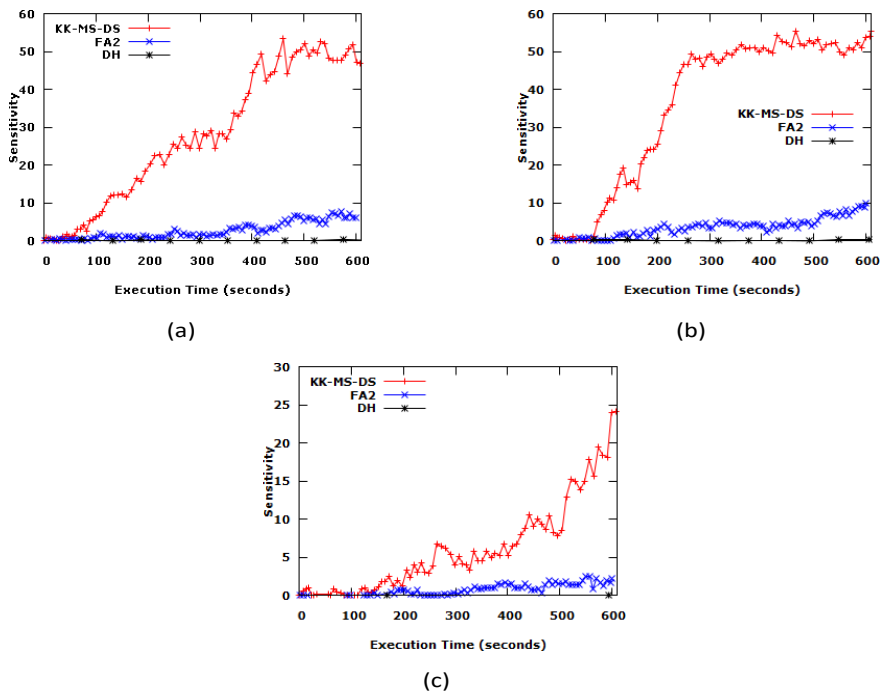

Fig. 14: Sensitivity of hole detection of FD algorithms with *n* = 3,000 and (a) *d* = 6, (b) *d* = 8, (c) *d* = 10 in WSNs.

*4.3.2 Dataset with 2,000 nodes*

Sensitivity of FD-CNN for different average degree is depicted in Figure 18. In this experiment, the total number of nodes is 2,000 and average degree 6, 8 and 10 are tested. The experimental results revealed that the proposed algorithm achieved the highest sensitivity when average degree *d* = 10 is used (see Figure 18(c)) for KK-MS-DS algorithm.

Specificity of FD-CNN for different average degree is depicted in Figure 19. In this experiment, the total number of nodes is set to 2,000 and average degree 6, 8 and 10 are tested. The proposed algorithm produced approximately 90% specificity when the average degree was high (i.e. *d* = 8 and *d* = 10, see Figure 19(b) and Figure 19(c)). the KK-MS-DS algorithm achieved the best result among three algorithms.

*4.3.3 Dataset with 3,000 nodes*

Sensitivity of FD-CNN for different average degree is depicted in Figure 20. In this experiment, the total number of nodes is set to 3,000 and average degree 6, 8 and 10 are tested. The experimental results revealed that sensitivity of the proposed algorithm algorithm was low (less than 20%) regardless of the average degree. From the experiment results, we found that when high average degree was used in

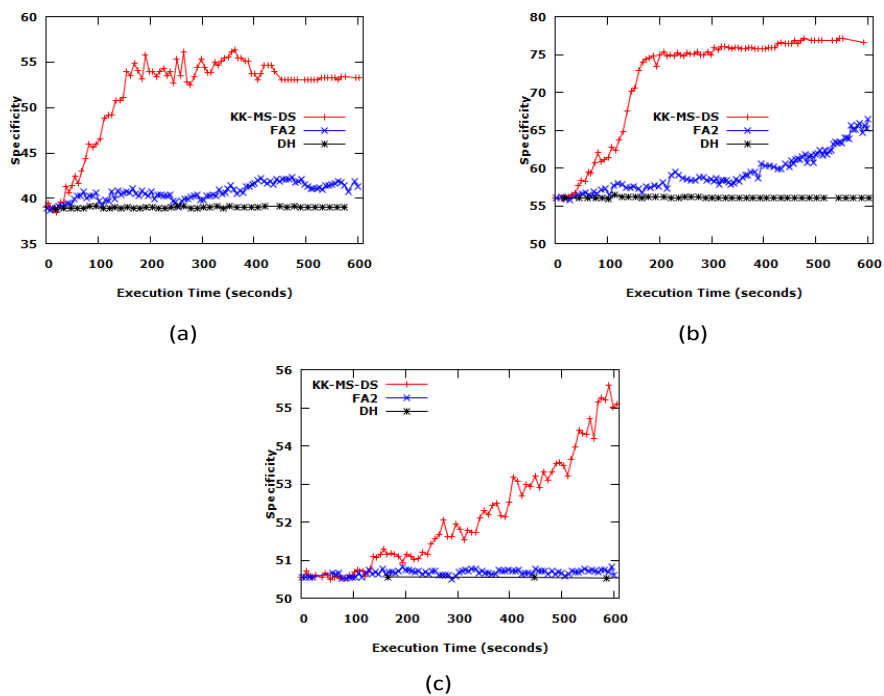

Fig. 15: Specificity of hole detection of FD algorithms with $n = 3,000$ and (a) $d = 6$, (b) $d = 8$, (c) $d = 10$ in WSNs.

dataset with 3,000 nodes, the generated layout from FD algorithms contain too many edge crossings

Specificity of FD-CNN for different average degree is depicted in Figure 21. In this experiment for uniform WSNs, the total number of nodes is set to 3,000 and average degree 6, 8 and 10 are tested. The proposed algorithm produced high specificity (approximately 87%) when the average degree was high (i.e. $d = 8$ and $d = 10$, see Figure 21(b) and Figure 21(c)).

*4.3.4 Summary*

According to the experimental results, the detection for KK-MS-DS and DH algorithms stops before time limit (i.e. 600 seconds) when $n = 1000$ since the node IDs located on the boundary of holes was unchanged for more than 10 iterations as illustrated in Figure 16. The experimental results revealed that the proposed FD-CNN achieved high specificity when $d = 10$. From the experiment results, we also found that the sensitivity of the FD-CNN from the layout generated by KK-MS-DS, FA2 and DH algorithms in uniform sensor networks are low regardless of the average degree and the total number of nodes.

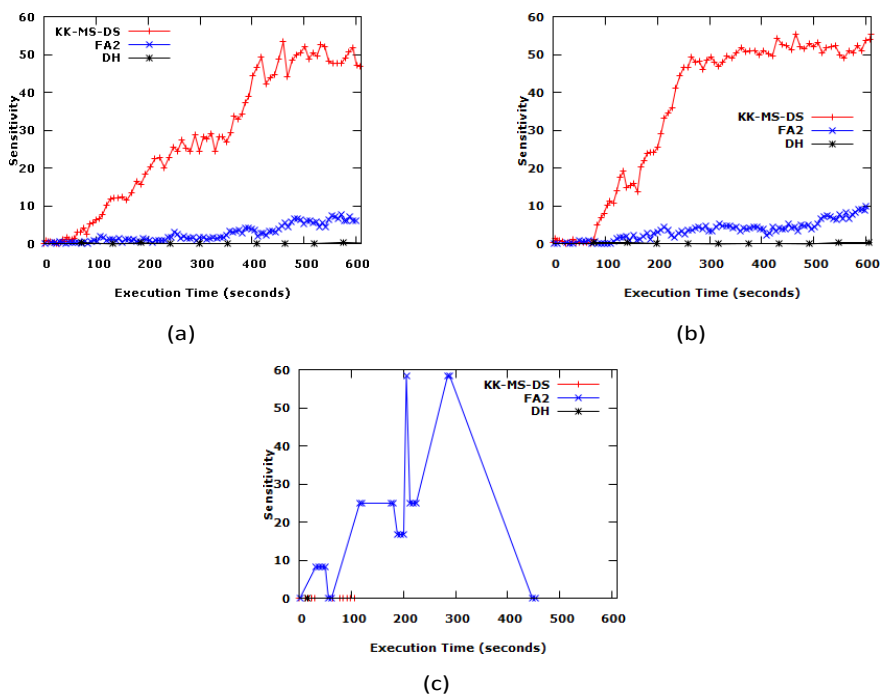

Fig. 16: Sensitivity when $n = 1,000$ and (a) $d = 6$, (b) $d = 8$, (c) $d = 10$ in uniform sensor networks.

4.4 Visualization of experiment results

Experimental results reveal that the performance of FD-CNN can vary significantly for different types of WSNs even with the same node count and average degree were used. To better understand the effect of FD algorithms in hole detection, we select some of the layouts produced by FD algorithms from the experiment results for visualization.

*Case 1:* Visualization of a layout which results high sensitivity during the experiments is shown in Figure 22. In this visualization, we can observe that the layout shown in Figure 22(b) which was generated by KK-MS-DS algorithm is similar to the ground truth shown in Figure 22(a). In this particular case, FD-CNN algorithm achieved approximately 80% sensitivity from the layout produced by KK-MS-DS algorithm. Moreover, we can also observe that the layout produced by the FA2 algorithm is twisted (see Figure 22(c)). DH algorithm also produced a layout which is also not similar to the ground truth and thus resulting a low sensitivity.

*Case 2:* The visualization of the layouts which result low sensitivity in the experiments is shown in Figure 23. From these visualizations, we can observe that the layout produced by three FD algorithms are both twisted and folded.

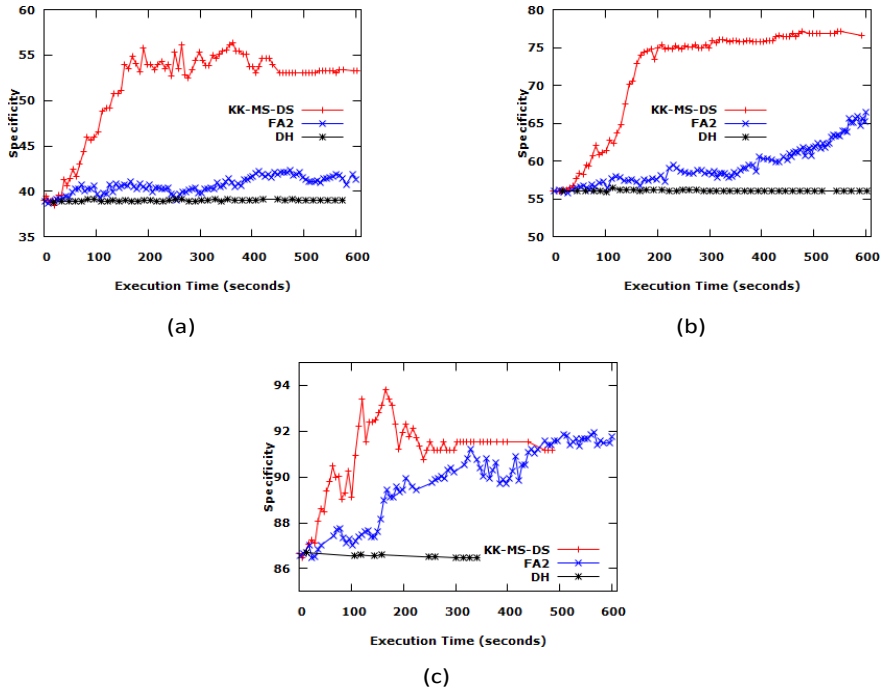

Fig. 17: Specificity when $n = 1,000$ and (a) $d = 6$, (b) $d = 8$, (c) $d = 10$ in uniform sensor networks.

*Case 3:* From the experimental results, we found that layouts in uniform WSNs achieved poor results. To better understand the underlying problems, we capture some of the layouts from the experiments and they are shown in Figure 24. From Figure 24(d), we can observe that KK-MS-DS algorithm can produce the most similar layout when it is compared to the ground truth (see Figure 24(a)). However, sensitivity result of this case is low since nodes are uniformly distributed and CNN cannot accurately recognize all the regions surrounded by the nodes.

## 5 Conclusion

In this paper, we propose a novel hole detection approach called FD-CNN for WSNs. The proposed approach takes advantage of the capability of Force-directed (FD) algorithms in generating potential layouts of the network and image/object detection ability of Convolutional Neural Networks (CNN). One of the main advantages of the proposed approach is its ability in detecting holes from WSNs without relying on any position information (i.e. $x$ and $y$ coordinates) of the nodes or location services such as GPS. The proposed approach is also able to identify the regions and nodes located on the boundary of the detected holes.

In our approach, a FD algorithm was deployed to produce the potential layouts from the topology of the WSN. The layout is then fed into the CNN model for hole detection. The detection process is iteratively executed until a predefined stopping criteria is satisfied. To generate labeled datasets for training CNN models, a pre-processing method was proposed. The proposed method includes 3 steps for preparing the labeled holes. The resulting labeled datasets are then used for training the CNN model.

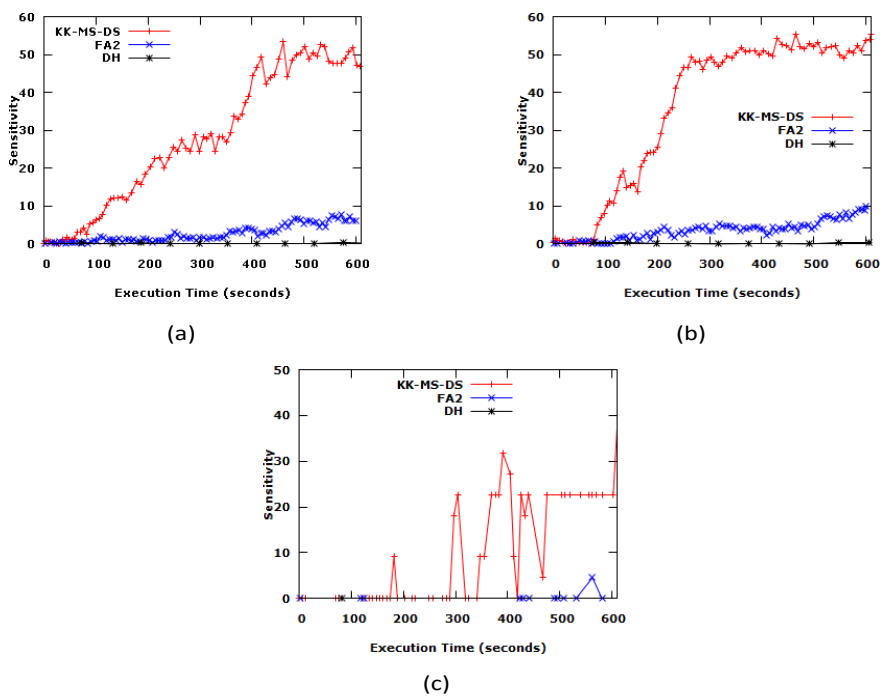

Fig. 18: Sensitivity when *n* = 2, 000 and (a) *d* = 6, (b) *d* = 8, (c) *d* = 10 in uniform sensor networks.

In the experiment section, we evaluate the FD-CNN by using the layouts generated from three FD algorithms. Experimental results show that FD-CNN can achieve 80% sensitivity and 93% specificity in less than 2 minutes. Experimental results also reveal that the performance of FD-CNN algorithm in WSNs are highly correlated to the quality of layouts generated by the FD algorithms. Average sensitivity of FD-CNN algorithm from the layout generated by KK-MS-DS algorithm is about 2.6 to 8 times higher than the layout generated by the FA2 and DH algorithms in WSNs.

However, performance of FD-CNN could be affected by several factors. First, in the proposed approach, a layout of the WSN is generated by using a FD algorithm. Therefore, positions of the sensors and distance among them in the layout are just the estimations projected by the FD algorithm and they could be different from the actual situation. Since, estimated layouts are used for hole detection, the sensitivity and specificity could be highly related to the performance of the FD algorithm used for estimation. Second, because the estimated payout of the WSN is converted into a bitmap image for processing, the scaling of the image can be complex, especially for large WSNs. Besides, the performance of hole detection by using the CNN models could be affected if there are large number of edge crossings in the bitmap image.

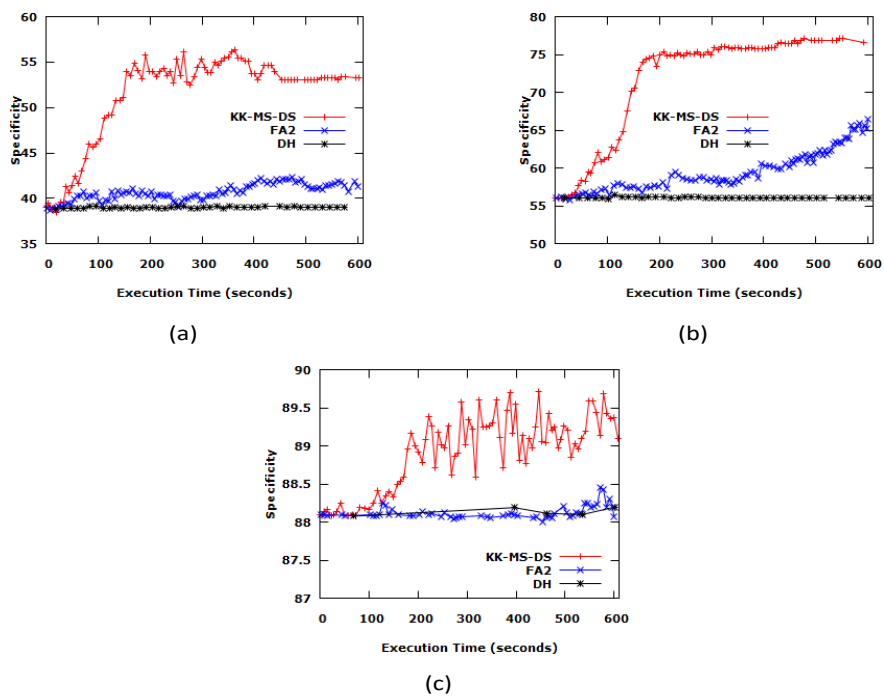

Fig. 19: Specificity when *n* = 2,000 and (a) *d* = 6, (b) *d* = 8, (c) *d* = 10 in uniform sensor networks.

As for the future work, we are planning to adopt force directed models to estimate the geographical locations of coverage holes and sensors in WSNs. To estimate the locations, FD algorithms could be combined with various metaheuristic algorithms. Moreover, we are also planning to develop a flexible scaling method to generate the datasets. Such scaling can be helpful in handling large WSNs. By using the datasets from flexible scaling method, we could minimize the complexity in training CNN models. In other words, re-training of the models from scratch could be avoided when the dimension and scale of bitmaps image is changed.

**Declarations**

This research was funded by University of Macau (File no. MYRG2019-00136-FST and MYRG2018-00246-FST).

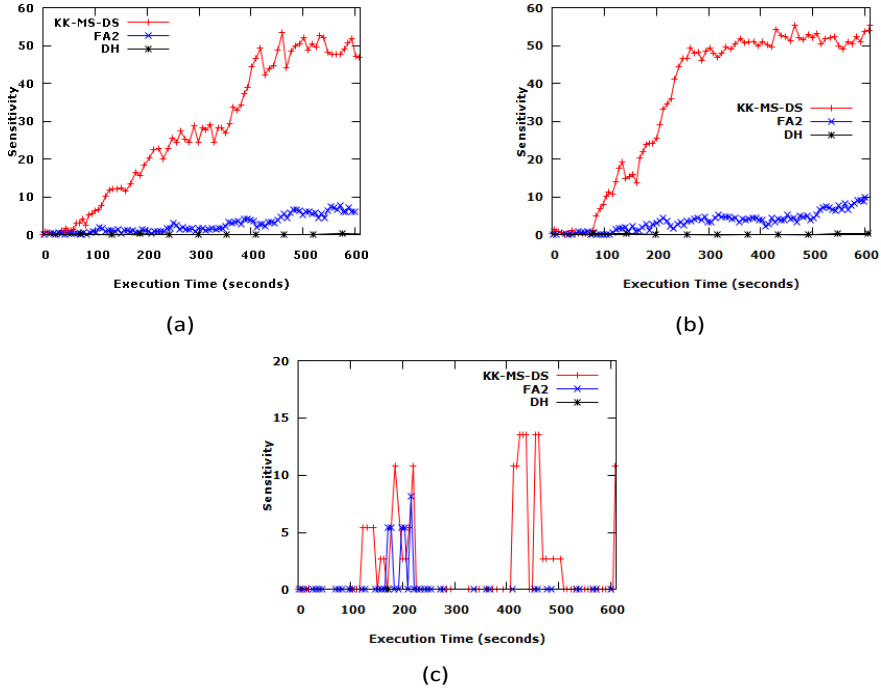

Fig. 20: Sensitivity of hole detection of FD algorithms with $n = 3,000$ and (a) $d = 6$, (b) $d = 8$, (c) $d = 10$ in uniform sensor networks.

## Appendix A - Force-directed Algorithms

*Davidson-Harel algorithm* The energy value $E$ used in the simulated annealing defined in the DH algorithm is the sum of attractive force ($f_a$) and repulsive force ($f_r$), which can be calculated as follows:

$$E = \sum_{i=1}^{n-1} \sum_{j=i+1} f_a(\|u_i - u_j\|) + f_r(\|u_i - u_j\|) \quad (1)$$

During initialization, a node $u$ is randomly selected from the network. Next, the DH algorithm creates a temporary node $v$. The DH algorithm then assigns the location to the node $v$ based on the location of the node $u$. The position of the node $v$ and other nodes in the network can be used to calculate the new energy value $E'$, which is defined as follows:

$$E' = \sum_{u \in V, v \notin V} f_a(\|u - v\|) + f_r(\|u - v\|) \quad (2)$$

In addition, when the liquid cools slowly, the DH algorithm obeys the Boltzmann distribution rule [14]. If $E' - E \leq 0$, then use $E'$ as the energy of the next iteration, because $E'$ has a lower energy value. If $E' - E > 0$, the probability equation is used to determine whether to use the new energy $E'$ in the next iteration. The probability equation is defined as follows:

$$p = e^{-\frac{E' - E}{k \times T}} \tag{3}$$

where $T$ is the temperature variable and $k$ is the Boltzmann constant. If the probability $p$ is less than the threshold $\varepsilon$, the new energy $E'$ is accepted; otherwise, the old energy $E$ will be used in the next iteration. The time complexity of DH algorithm is $O(V^2 \times E)$, where *V* is the number of nodes in the network topology, *E* is the number of edges in the network topology.

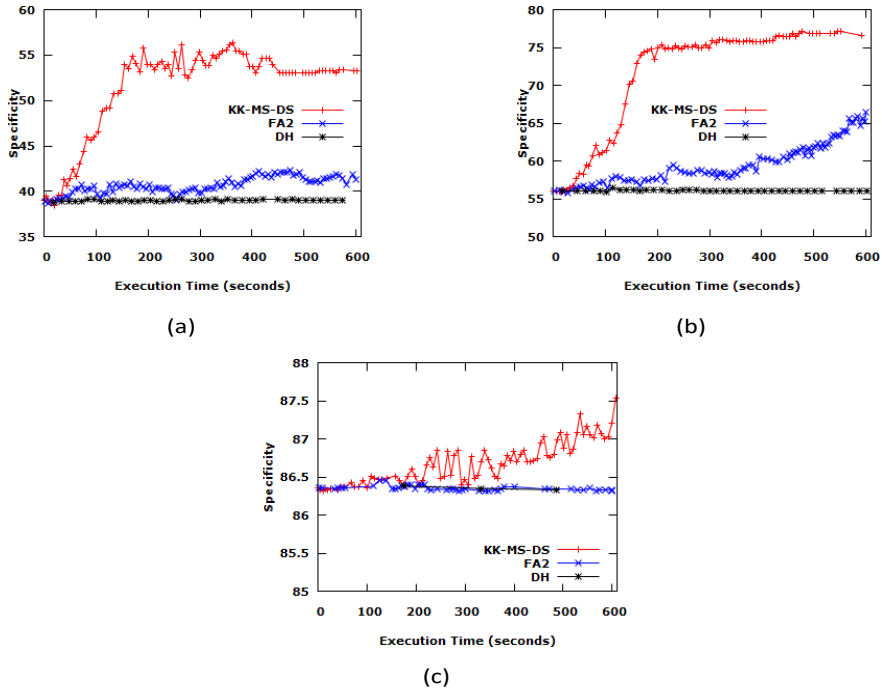

Fig. 21: Specificity of hole detection of FD algorithms with $n = 3,000$ and (a) $d = 6$, (b) $d = 8$, (c) $d = 10$ in uniform sensor networks.

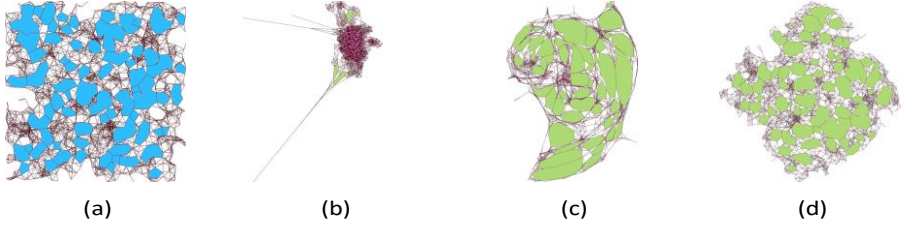

Fig. 22: Layouts of sparse WSN with $n = 1,000$ and $d = 8$. (a) ground truth, (b) layout generated by the DH algorithm, (c) layout generated by the FA2 algorithm, and (d) layout generated by the KK-MS-DS algorithm.

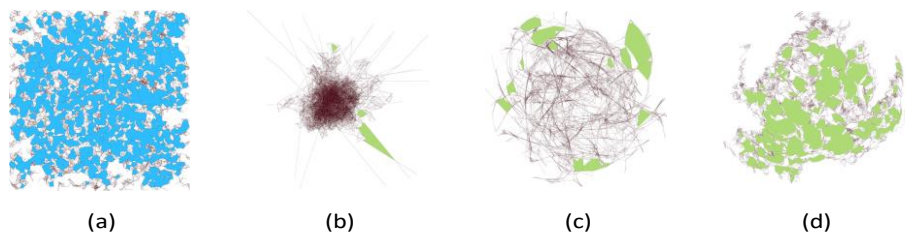

(a)  (b)  (c)  (d)

Fig. 23: Layouts of sparse WSN with *n* = 3, 000 and *d* = 6. (a) ground truth, (b) layout generated by the DH algorithm, (c) layout generated by the FA2 algorithm, and (d) layout generated by the KK-MS-DS algorithm.

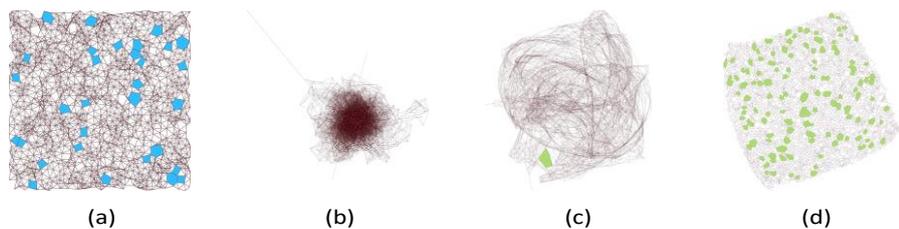

(a)  (b)  (c)  (d)

Fig. 24: Layouts of uniform WSN with *n* = 1, 000 and *d* = 8. (a) ground truth, (b) layout generated by the DH algorithm, (c) layout generated by the FA2 algorithm and (d) layout generated by the KK-MS-DS algorithm.

*ForceAtlas2 algorithm* Jacomy et al. [20] proposed a revised attractiveness based on the LinLog model, which is defined as follows:

$$F_a(u, v) = log(1 + d(u, v)) \qquad (4)$$

where *d* is the distance between nodes *u* and *v*. In addition, a degree-dependent repulsion model is proposed in the FA2 algorithm to reduce the repulsive force. This repulsion model increases the chance that nodes with lower than average degrees are connected to nodes with higher than average degrees. The repulsion model of FA2 algorithm is defined as follows:

$$F_r(u, v) = k \frac{(deg(u) + 1) \times (deg(v) + 1)}{d(u, v)} \qquad (5)$$

where *k* is the ideal pairwise distance constant, as used in FR algorithm. *d* is the distance between nodes *u* and *v*, and *deg(n)* is the number of associated edges Node *n*, including the edge of in-out degree. The time complexity of FA2 algorithm is $O(V^2 + E)$ where *V* is the number of nodes in the network topology, *E* is the number of edges in the network topology.

*KK-MS-DS algorithm* The goal of the KK-MS-DS algorithm [8] is to push the nodes in the outer boundary away from the inner nodes. Nodes tag with a decaying stiffness ($m$). The higher the decaying stiffness value of the node ($m$), the farther the distance of the node can be moved. The value of $m$ of the node decreases with the execution time, which is defined as follows:

$$m' = m - zp^t \qquad (6)$$

where $t$ is the number of times that the selected node is updated. $p$ is the decay rate, and $z$ is the remaining energy possessed by the node. The KK-MS-DS algorithm terminates, when the stable state ($r$) remains unchanged until a predefined iteration or $r$ is less than the threshold $\varepsilon$. The stable state ($r$) means that a coarse visualization of the graph has been constructed, but the final stage of the entire graph has not yet been reached. The ratio of the stable state $r$ is defined as follows:

$$r = \frac{\frac{1}{d}\sum_{i=1}^{d}|L_i' - L_i|}{\sum_{i=1}^{d} L_i - \frac{1}{d}\sum_{i=1}^{d} L_i - L_i} \qquad (7)$$

where $d$ is the total number of edges in the graph, $L_i'$ is the edge length from the graph generated by the KK-MS-DS algorithm, and $L_i$ is the edge length of the input graph. The time complexity of KK-MS-DS algorithm is $O(n \times (V + v^2))$, where $n$ is the number of iteration, $V$ is the number of nodes in the given graph, and $v$ is the number of nodes in the ordered queue.

**Appendix B - Pseudocode of Force-directed Algorithms**

The pseudocode for DH algorithm, FA2 algorithm and KK-MS-DS algorithm are given in Algorithm 4, Algorithm 5 and Algorithm 6 respectively.

**ALGORITHM 4:** Pseudocode of DH algorithm.

**Input:** Network topology $G = (V, E)$.
**Output:** A layout of network topology of $G$.
initialize radius
  $r = max(\frac{RADIUS-WEIGHT-CONSTANT}{5}, \frac{max(v.x)-min(v.x), max(v.y)-min(v.y)}{5})$ for
  $v \in V$;
initialize iteration $it$;
// Initialize Energy $E$
**for** $u, v \in V, u \mathrel{/}= v$ **do**
    $E = E + \frac{1}{\|u-v\|} + \|u - v\|$;
**end**
**while** $it > 0$ **do**
    // Compute Candidate Layout
    initialize a network topology $G' = (V', E')$;
    $G' = G$;
    **while** $t > \epsilon$ **do**
        select a temporary node $v$ from $G'$;
        // The positioin of node $v \in V'$ is based on the position of $r \in V$
        Select a node $r$ from $V$ randomly;
        $v.x = r.x + rand(0, r)$;
        $v.y = r.x + rand(0, r)$;
        // Assign random angle to node $v$
        $v.x = Math.cos(rand(0, 2 * PI)) * r$;
        $v.y = Math.sin(rand(0, 2 * PI)) * r$;
        $t = t - 1$;
    **end**
    // Compute Candidate Energ $E'$ and Test the Energy of Candidate Layout $G'$
    **for** $u, v \in V', u \mathrel{/}= v$ **do**
        $E' = E' + \frac{1}{\|u-v\|}$;
        $E' = E' + \|u - v\|$;
    **end**
    **if** $E' < E$ **then**
        $G = G'$;
    **end**
    // Update iteration
    $r = r \times SHRINK - CONSTANT$;
    $it = it - 1$;
**end**

**ALGORITHM 5:** Pseudocode of FA2 algorithm.

**Input:** Network topology $G = (V, E)$.
**Output:** A layout of network topology of $G$.
initialize $v_{mass} = 0$ for $v \in V$;
initialize iteration $it$;
**while** $it > 0$ **do**
    // Compute Repulsion Force
    **for** $u, v \in V_{WT}, u = v$ **do**
        $r = R * \frac{u_{mass} \times v_{mass}}{\|u-v\|^2}$;
        $u_{dx} = u_{dx} + \|u - v\| \times r$; $u_{dy} = u_{dy} + \|u - v\| \times r$;
        $v_{dx} = v_{dx} - \|u - v\| \times r$; $u_{dy} = u_{dy} - \|u - v\| \times r$;
    **end**

    // Compute Graviation Force
    **for** $u \in V$ **do**
        $g = S \times \frac{u_{mass} \times G}{\|u-v\|^2}$;
        $u_{dx} = u_{dx} + \|u - v\| \times g$; $u_{dy} = u_{dy} + \|u - v\| \times g$;
    **end**

    // Compute Attraction Force
    **for** $e \in E$ and $u, v$ are start and end nodes of the $e$ **do**
        $a = -S \times A$;
        $u_{dx} = u_{dx} + \|u - v\| \times a$; $u_{dy} = u_{dy} + \|u - v\| \times a$;
        $v_{dx} = v_{dx} - \|u - v\| \times a$; $u_{dy} = u_{dy} - \|u - v\| \times a$;
    **end**

    // Compute Swinging and Target Speed for $v$ in $V$
    // Apply forces to nodes
    **for** $u \in V$ **do**
        $f = \frac{S}{1+\sqrt{targetspeed+swinging}}$;
        $u_x = u_{dx} + \|u - v\| \times f$; $u_y = u_{dy} + \|u - v\| \times f$;
    **end**

    // Update iteration
    $it = it - 1$;
**end**

---

**ALGORITHM 6:** Pseudocode of KK-MS-DS algorithm.

---

**Input:** Network topology $G = (V, E)$.
**Output:** A layout of network topology of $G$.
initialize the start area $WT = (W_{WT}, E_{WT})$ that $WT \subseteq T$;
initialize the iteration count $it = 0$;
initialize the starting node $s$ which has a maximum average degree in $G$;
// Step 1.
add node $s$ into $V_{WT}$;
**for** $v \in V$ **do**
    initialize $v_m$ and add node $v$ into $V_{WT}$ where $hopcount(V_{WT}, v) = 2$;
**end**
// Step 2 & 3.
**while** $WT = T$ **do**
    **if** $r < \epsilon$ **then**
        **for** $v \notin V_{WT}, v \in V$ **do**
            add node $v$ into $V_{WT}$ where $hopcount(V_{WT}, v) = 2$;
            $V_m = \frac{K}{d_{i,j}^2}$;
        **end**
        **for** $u, v \in V_{WT}, u \neq v$ **do**
            update $L_i$ for node $v$ and $u$;
        **end**
    **else**
        $G = KK - MS(WT, 5\%)$;
        update $L_i$ for node $v$ in $V_{WT}$;
    **end**
**end**
// Step 4.
clear $V_{WT}$ and $E_{WT}$ in $WT$;
**for** $v \in V$ **do**
    add node $v$ into $V_{WT}$;
    $v_m = \frac{K}{d_{i,j}^2}$;
**end**
compute $r$ for the $WT$;
**while** $r \geq \epsilon$ **do**
    $G = KK - MS(WT, 5\%)$;
**end**

---